# Designs for a Quantum Electron Microscope

*P. Kruit[1*], R. G. Hobbs[2], C-S. Kim[2], Y. Yang[2], V. R. Manfrinato[2], J. Hammer[3], S. Thomas[3], P. Weber[3], B. Klopfer[4], C. Kohstall[4], T. Juffmann[4], M. A. Kasevich[4], P. Hommelhoff[3], K. K. Berggren[2].*

[1]Department of Imaging Physics, Delft University of Technology, Lorentzweg 1, 2628CJ Delft, The Netherlands.
[2]Research Laboratory of Electronics, Massachusetts Institute of Technology, Cambridge, Massachusetts, 02139, United States.
[3]Department of Physics, Friedrich Alexander University Erlangen-Nürnberg (FAU), Staudtstrasse 1, D-91058 Erlangen, Germany.
[4]Department of Physics, Stanford University, Stanford, California, 94305, United States.

[*]p.kruit@tudelft.nl

**Abstract**

One of the astounding consequences of quantum mechanics is that it allows the detection of a target using an incident probe, with only a low probability of interaction of the probe and the target. This 'quantum weirdness' could be applied in the field of electron microscopy to generate images of beam-sensitive specimens with substantially reduced damage to the specimen. A reduction of beam-induced damage to specimens is especially of great importance if it can enable imaging of biological specimens with atomic resolution. Following a recent suggestion that interaction-free measurements are possible with electrons, we now analyze the difficulties of actually building an atomic resolution interaction-free electron microscope, or "quantum electron microscope". A quantum electron microscope would require a number of unique components not found in conventional transmission electron microscopes. These components include a coherent electron-beam splitter or two-state-coupler, and a resonator structure to allow each electron to interrogate the specimen multiple times, thus supporting high success probabilities for interaction-free detection of the specimen. Different system designs are presented here, which are based on four different choices of two-state-couplers: a thin crystal, a grating mirror, a standing light wave and an electro-dynamical pseudopotential. Challenges for the detailed electron optical design are identified as future directions for development. While it is concluded that it should be possible to build an atomic resolution quantum electron microscope, we have also identified a number of hurdles to the development of such a microscope and further theoretical investigations that will be required to enable a complete interpretation of the images produced by such a microscope.

**Introduction**

Electron microscopy (EM) has revolutionized our understanding of biomolecules, cells, and biomaterials, by enabling their analysis through imaging with (near-) atomic-scale resolution.[1–3] Imaging biological systems at such high resolution has facilitated the determination of structure-property correlations that are key to our understanding of the biological function of such systems. However, the high-energy electrons typically used for electron microscopy are known to cause damage to biological specimens due to inelastic scattering events that break chemical bonds (radiolysis), heat the specimen, and even dislocate atoms (knock-on damage).[4] Specimen damage is related to the fact that image information is shot-noise limited, meaning that a minimum number of electrons $N_e$ is required to form an image with a specified signal-to-noise ratio. Atomic-scale imaging of biological samples requires a lower bound $N_e$, which is already more than sufficient to cause damage. Consequently, resolution must be sacrificed in imaging biological specimens to avoid this damage. Presently, the attainable resolution for single objects by cryo-electron tomography is approximately 5-10 nm, which prevents the production of atomically resolved images.[5] Sub-nanometer resolution is regularly achieved in cryo-electron tomography through the use of sampling, where thousands of images with different projections of identical macromolecules are acquired and combined to generate the 3-D structure computationally.[6–8] However, this approach adds the requirement for the acquisition of thousands of images of identical structures as well as significant computational effort. Moreover, the use of sampling techniques does not address the fundamental problem, which is that a route toward atomically resolved images of individual, isolated biological specimens, with negligible damage to the sample, is still lacking.

Recent advances in quantum metrology[9] might allow us to overcome the resolution limits imposed by shot-noise. Instead, the Heisenberg uncertainty principle limits resolution so that the accuracy of a measurement scales as $1/N_e$, a $\sqrt{N_e}$ improvement over the shot-noise limit. Recently, approaches implementing measurement techniques based on quantum mechanical phenomena have been proposed that could reduce the electron dose to the specimen in a transmission electron microscope (TEM).[10–12] Here, we follow up on another idea first reported by Elitzur and Vaidman,[13] who proposed that an opaque object may be observed by detecting a photon that did not interact with that object.[14] This concept, shown schematically in Fig. 1, uses a Mach-Zehnder interferometer to perform an "interaction-free measurement" (IFM). Here, a fully opaque object in the lower arm of the interferometer destroys interference in the system, and leads to the detection of a photon at the dark output of the interferometer (D2 in Fig. 1). Earlier concepts for such 'negative result' experiments were proposed separately by Renninger and Dicke,[15,16] but the designs considered in our work stem from the Elitzur and Vaidman scheme. While the experiment proposed by Elitzur and Vaidman may achieve a truly interaction-free measurement for an ideal, fully opaque specimen, any measurement of a partially transparent specimen or a quantum system is not likely to fulfill the requirements of an IFM and so we refer to such measurements more generally here as quantum interrogations.[17] Various realizations of such quantum interrogations have been achieved with photons and shall be discussed later. Here, we show how this idea could be implemented in an electron microscope, not dissimilar to existing scanning transmission electron microscopes, to yield the structure of a specimen with negligible exposure of the specimen, and thus reduced damage to the specimen. We will refer to such a microscope as a 'Quantum Electron Microscope' (QEM), referring to the essential quantum mechanical aspect of the interrogation technique.

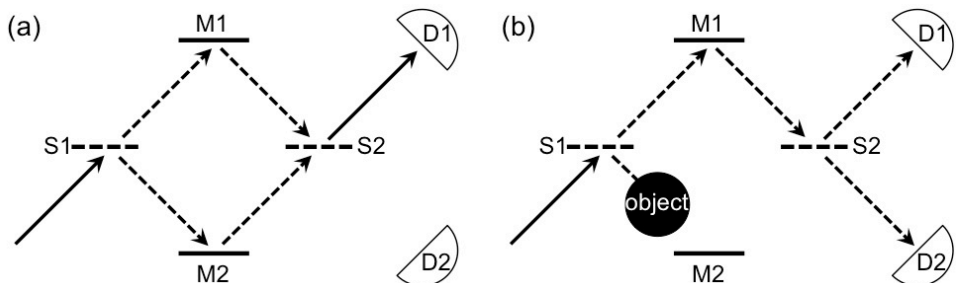

**Fig. 1.** An interaction-free measurement based on a Mach-Zehnder interferometer. (a) An incident particle (photon/electron) enters the interferometer from the left and is split into a superposition of two wave-packets by a 50/50 beam-splitter (S1). The two wave-packets then travel through the two arms of the interferometer via mirrors M1, and M2 respectively, before entering a second 50/50 beam-splitter (S2). The interferometer geometry is arranged such that a particle is never detected at detector 2 (D2) due to deconstructive interference, while a particle is always detected at detector 1 (D1). (b) When an object is placed in the path of the lower arm of the interferometer, deconstructive interference of the two-wave-packets is prevented, and a particle may now be detected at D2. In that case, the presence of the object has been determined by detection of a particle that did not interact with the object.

Fig. 1 illustrates the IFM scheme proposed by Elitzur and Vaidman[13] based on a Mach-Zehnder interferometer. A probe particle/wave, in this case a photon, is coherently split into two beams of equal amplitude (S1), a sample beam and a reference beam. The two paths are recombined at a second 50/50 beam-splitter (S2) and, if the relative path length is chosen correctly, interference means that the photon will always exit this beam splitter at the same output port (D1). However, when an object blocks the path of the sample-beam, a photon can reach the other output port (D2) with a probability of ¼. In this case, the presence of the object was detected, although there was no interaction between the object and the detected photon. This technique was conceptually extended to success probabilities arbitrarily close to one using an approach analogous to a discrete form of the quantum Zeno effect.[18–21] The approach reported by Kwiat *et al*.[18] involved increasing the reflectivity of the beam-splitters, as Elitzur-Vaidman had previously suggested, thus reducing the intensity of the beam directed at the specimen of interest, while extending the system to probe the specimen multiple times within a single measurement. Kwiat *et al*.[18] proposed the use of a cavity to enable the coherent evolution of intensity transfer between the sample and reference beams to achieve higher detection probabilities while simultaneously reducing the intensity at the specimen. These higher detection probabilities have been experimentally demonstrated[18,22,23] and subsequently applied to imaging systems.[24]

The coherent evolution of intensity transfer between the sample and reference beams in the scheme described by Kwiat *et al*. may be represented by a quantum mechanical two-level system such as a shallow electronic double well potential (Fig. 2(a)). Thus, the scheme proposed by Kwiat *et al*. may be applicable to electrons as well as photons, as described by Putnam and Yanik.[25] The two lowest energy eigenstates of a double-well potential are the symmetric state $|\Psi_S\rangle$ and the antisymmetric state $|\Psi_A\rangle$, as shown in Fig. 2(a).

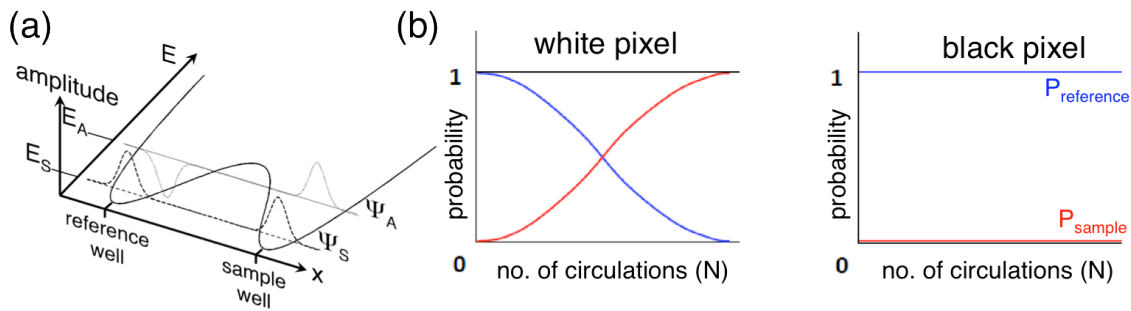

**Fig. 2.** (a) Double well potential. The two lowest energy eigenstates of this potential well are the symmetric state labeled $|\Psi_S\rangle$ (dashed line) and the antisymmetric state labeled $|\Psi_A\rangle$. (b) Plots of the probability of finding an electron in the reference and sample rings (see text), which contain the reference and sample wells respectively, as a function of the number of circulations in those rings. The plots shown are for the case where an opaque object is absent from the sample ring (white pixel), or is present in the sample ring (black pixel).

The state of an electron localized in the left (right) well of the double-well system can be defined as $|reference\rangle$ ($|sample\rangle$), which can be represented as a linear combination of the eigenstates: $|reference\rangle = \frac{1}{\sqrt{2}}(|\Psi_S\rangle - |\Psi_A\rangle)$, $|sample\rangle = \frac{1}{\sqrt{2}}(|\Psi_S\rangle + |\Psi_A\rangle)$. The electron starts

in the reference well and hence the initial state is the localized state |*reference*⟩. Since this state is not an eigenstate, the state of the electron will change with time. According to the dynamics of a two-level system, the state will gradually evolve back and forth between states |*reference*⟩ and |*sample*⟩. Specifically, the electron population will gradually transfer back and forth between the two wells. If we consider the double-well potential as being generated by a pair of coupled electron ring-resonators, then the electron population will move between the reference well, which is localized on the reference ring, and the sample well, which is localized on the sample ring, with each circulation in those rings as shown in the left plot of Fig. 2(b).[26] However, if an opaque object is placed in the path of the sample ring, it will destroy the coherent evolution of the electron state, and the electron will remain in the reference ring as shown in the plot on the right in Fig. 2(b), analogous to the quantum Zeno effect.[19] As a result, determining the well/ring occupied by the electron will reveal whether a transparent or opaque object exists in the path of the sample ring. A full binary image of the specimen may be obtained pixel-by-pixel, after $N$ circulations, by scanning the sample ring across the specimen, similar to a scanning electron microscope (SEM) or a scanning transmission electron microscope (STEM).

Recently, Putnam and Yanik[25] outlined a way to implement this measurement principle using electrons. Their conceptual approach is based on using a double-well potential, where continuous coupling of the electron states is achieved by quantum mechanical tunneling from one well to the other. They suggest an implementation of the approach where the double potential well is created in a ring-shaped hybrid Paul Penning trap for electrons. A sample could be placed in one of the rings, as shown in Fig. 3.

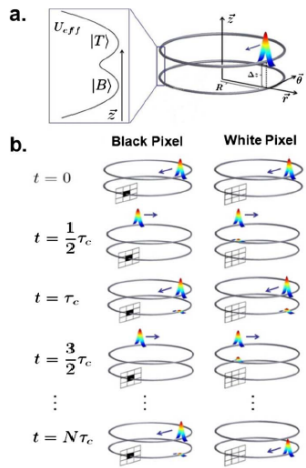

**Fig. 3.** (a) Continuously coupled electron ring guides proposed by Putnam and Yanik. The figure shows an electron wave-packet circulating in the upper ring and the inset shows $U_{\text{eff}}$ the double-well potential that defines the electron rings labeled according to the states that occupy each, which are labeled |T⟩ and |B⟩ in the figure, but shall be referred to as |*reference*⟩ and |*sample*⟩ respectively in this text. (b) A schematic description of the imaging mechanism. The grid in the path of the lower electron ring represents the object to be imaged, which consists of both transparent and opaque pixels thus creating a binary image. A black (white) pixel in the path of the lower ring prevents (allows) coherent evolution of the electron wave-packet from the upper ring to the lower ring after $N$ circulations. Reprinted with permission from Putnam and Yanik.[25] Copyright 2009 The American Physical Society.

We shall now consider the issue of damage to the sample during the imaging procedure. Radiation damage can be shown to scale with $1/N_o$,[27] where $N_o$ is the number of circulations by the electron in the ring required to fully tunnel from one ring to the other *i.e.* for a complete transition from state |*reference*⟩ to state |*sample*⟩. The origin of the relationship between damage and $N_o$ may be understood by considering the process of transferring the

electron wavefunction from the |*reference*⟩ state to the |*sample*⟩ as the transfer of $1/N_o$ of the *amplitude* from the reference beam to the sample beam during each circulation. When a sample is present and interacts with the wavefunction in the |*sample*⟩ state, the damage per circulation is then proportional to the *intensity* in the sample ring, that being $1/N_o^2$. Consequently, the total damage after $N_o$ circulations will be proportional to $1/N_o$. Thus, by making $N_o$ arbitrarily large, a quantum measurement can in principle be achieved with vanishing damage probability.

The description of the QEM as given above outlines a route toward binary, *black and white* imaging, which may find use in a variety of biological imaging applications such as high-contrast stained objects or materials infused with nanoparticles. One problem with the interaction-free scheme of detection however, is that it is not immediately clear how one can obtain grey values in an image, which is desirable in electron microscopy. A recent paper from a number of the authors of this work concluded that grayscale imaging in the interaction-free scheme as described above is not possible without inducing damage to the sample, albeit less damage than in conventional schemes for high-contrast specimen.[27] However, Facchi *et al*.[28] have suggested that grayscale imaging may be possible for largely opaque objects using an imaging scheme referred to as quantum Zeno tomography. Another problem, or perhaps opportunity, is the fact that an electron microscope sample can also change the phase of the sample beam, which influences the subsequent coupling between the reference beam and the sample beam. The issue of the electron phase-shift in a QEM system was considered by Thomas et al.,[27] but has not yet been analyzed in detail. In general, the field of interaction-free electron microscopy is so novel that the possibility of grayscale imaging using such a technique has not yet been completely eliminated, and will be the subject of further investigation.

Putnam and Yanik[25] predicted an imaging resolution of 19 nm for their system consisting of a hybrid Paul-Penning trap using 100 keV electrons. Apart from the limited resolution, many details of the system described by Putnam and Yanik present challenges for the construction of a proof-of-concept QEM, for example, their system is based on charged particle trapping techniques, which are not commonly used in electron microscopy and are less suitable for high-energy electrons, which are typically required for imaging applications. In this work, we will describe a number of competing approaches toward the first demonstration of an operational QEM, with an emphasis upon the practical requirements for a successful design. In particular we will consider systems based on four different electron-beam couplers to fulfill the required coupling between the |*reference*⟩ and |*sample*⟩ states in the QEM. Here, we consider discrete coupler units as opposed to the continuous coupler unit described by Putnam and Yanik. Discrete coupler units will enable greater compatibility with conventional electron microscopy components, a key requirement for the successful development of a QEM system. Below, we highlight the key components required in a QEM.

A QEM must have the following sub-units:
1. A central electron-optical unit in which the electron beam can travel repeatedly through the two-state coupler, which creates the necessary coupling between the reference and the sample states. We call this the "resonator".
2. A two-state-coupler positioned within the resonator.
3. The resonator should also contain a sample holder and focusing unit to either create a small probe on the sample or to form a magnified image of the sample depending on whether the microscope is a scanning transmission microscope or an imaging transmission microscope.

4. An electron source and beam-forming unit.
5. A unit that opens the "resonator" to allow the electron beam to enter. We call this unit the entrance "barn door".
6. An exit "barn door".
7. A detector unit.

The modular design of the proposed QEM instrument is shown schematically in Fig. 4 below, and is applicable to all two-state-coupler systems considered in this work.

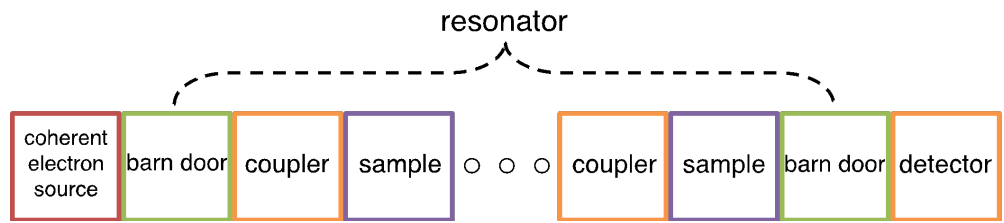

**Fig. 4.** Modular design of the QEM system showing the layout and position of the resonator. The modular framework will enable compatibility with existing electron microscopes, potentially allowing existing electron microscopes to operate in QEM-mode after installation of a resonator.

While the coherent electron source and the electron detector are standard components of TEMs, a resonator and a coherent two-state coupler are not. We have found that the design of an operational QEM is to a large extent determined by the choice of the two-state-coupler. So far we have identified four possible couplers: (1) a transmission grating such as a thin crystal, (2) an electron mirror patterned with a grating, (3) a radio frequency potential well, and (4) a standing light wave transmission grating. In the next section, we shall describe the preliminary designs of quantum electron microscopes based on each of these four two-state-couplers.

**1. Design based on a thin crystal coupler.**

Bragg diffraction from a crystalline sample can be used to coherently split a beam of electrons. In fact, crystalline electron beam-splitters were adopted in the first electron interferometry device developed by Marton in the 1950s, and have been implemented in a number of other electron interferometer devices since then,[29–33] (thin crystals have since been replaced by electron biprisms in electron interferometry instrumentation[34]). The dynamical theory of electron diffraction defines that the direct beam and the diffracted beams are coupled by the atomic potential of the crystal, and hence that the intensity in the beams oscillates as a function of the crystal thickness.[4,35,36] Commonly observed "thickness fringes" in TEM images are caused by this intensity oscillation between the direct beam and diffracted beams. Fig. 5 shows bright-field and dark-field TEM images of a crystalline Si wedge. Bright-field images are formed by collecting the direct-beam alone by using the objective aperture of the microscope to isolate the direct-beam (000 in Fig. 5(c)). The dark-field image shown here was formed by collecting one of the first-order diffracted beams. Thickness fringes can be observed at the edge of the wedge and are caused by the reducing thickness of the Si crystal toward the lower edge shown in Fig. 5. The thickness variation leads to intensity oscillations such that the intensity in the direct beam decreases as the intensity in the diffracted beams increases, thus leading to contrast variation in a bright-field TEM image where only the direct-beam is collected by the objective aperture to form an image (Fig. 5(a)). The inverse contrast can then be seen in the dark-field image. This intensity oscillation is known as "*Pendellösung*" after the German for "pendulum solution".[37]

Fig. 5(b) shows plots of the electron beam intensity for both the direct beam and the diffracted beam as a function of crystal thickness. The plots in Fig. 5(b) were produced by solving the Howie-Whelan equations for the two-beam condition (see Fig. 5(c)).[4] The periodic oscillation of the electron beam intensity in both the direct, and the diffracted beams can be considered analogous to the intensity oscillations in the reference and sample beams of a QEM as described in Fig. 3. For the purpose of the quantum interrogation in a QEM, we may replace the thick crystal described by dynamical diffraction by a very thin crystal, and lead the same electron through it many times. The oscillatory behavior should then be identical to that of a thick crystal if, by means of electron optics, we can reconstruct the electron wavefunction produced at the exit plane of the crystal, at the entrance plane of the crystal, after a single pass through the resonator.

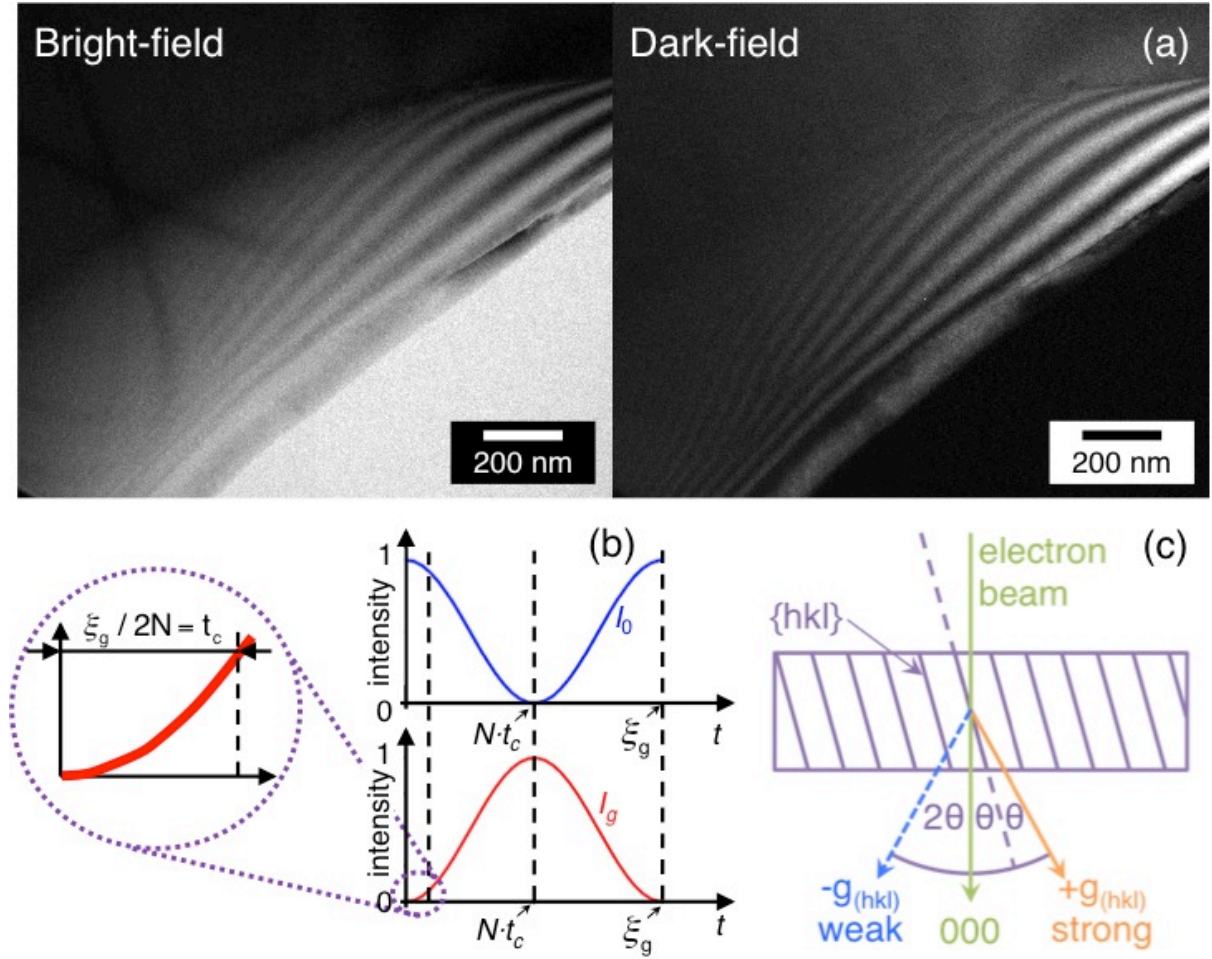

**Fig. 5.** (a) Bright-field and dark-field TEM images of a crystalline Si wedge acquired under the two-beam condition with a <110> oriented crystal and the direct beam and a {220} beam strongly excited. Thickness fringes can be observed terminating at the edge of the wedge to the lower right of each image. (b) Plots of electron beam intensity as a function of crystal thickness ($t$) for both the direct beam (reference beam in QEM) and the diffracted beam (sample beam in QEM) based on solutions to the Howie-Whelan equations,[4] which stipulate that $I_g = \left(\frac{\pi t}{\xi_g}\right)^2 \left(\frac{sin^2(\pi t s_{eff})}{(\pi t s_{eff})^2}\right) = 1 - I_0$  $s_{eff} = \sqrt{s^2 + \frac{1}{\xi_g^2}}$, where $I_g$ is the intensity of the diffracted beam, $I_0$ is the intensity in the direct beam, $t$ is sample thickness, $\xi_g$ is the extinction distance, and $s$ is a crystal-dependent parameter. If we consider a series of thin crystals with thickness $t_c$ in the path of the electron-beam, we can observe a full intensity transfer from the direct-beam to the diffracted beam after $N$ passes through that thin crystal. The plots in (b) show both the oscillatory nature of the intensity transfer between the direct and diffracted beams known as the "*Pendellösung*" effect, and the point at which complete intensity transfer occurs ($N \cdot t_c$), which would be the point of operation for QEM. (c) Schematic representation of a typical two-beam

diffraction condition for a thin crystalline sample. The figure shows the direct beam (000) and diffracted beams ($g_{(hkl)}$) for an incident electron beam at a {$hkl$} family of lattice planes that are oriented at the Bragg angle ($\theta$) with respect to the incident beam. In this geometry, only one strong diffracted beam is formed.

The calculated intensity oscillations shown in Fig. 5(b) were based on the assumption that a perfect two-beam condition was obtained where all of the electron-beam intensity is contained within the direct beam and a single diffracted beam. Moreover, we assumed that any losses in intensity due to inelastic scattering within the thin crystal coupler were negligible. Successful implementation of this coupler in a resonator for a QEM system will depend upon satisfactory solutions to the following issues: (1) intensity losses to other diffracted pathways should be eliminated or blocked without too much impact on the intensity contained within the reference and sample beams; (2) inelastic scattering at the crystal coupler needs to be minimized to reduce unwanted intensity loss; and (3) the relative phases of the sample and reference beams need to be correct when they complete a circulation in the resonator in order to achieve coherent coupling of the two beams in subsequent circulations. While the issues highlighted here are challenging to overcome, in the following section we propose routes to address each of these issues during the system design.

We propose to address issue (1) – that of losses to higher order diffractive pathways – by reducing the number of diffractive pathways available for the beam to exit the crystal. The number of diffractive pathways to exit the crystal coupler may be reduced by using a crystal with a zone axis consisting of closely spaced lattice planes. Crystals possessing zone axes with short interplanar distances (d-spacing) are desirable to increase the Bragg angle and thus reduce the number of diffracted beams falling on the Ewald sphere resulting in a cleaner two-beam condition. Additionally, we propose that an aperture placed near the sample plane would behave as an opaque pixel for any unwanted diffracted beams. As such, the aperture may function in a similar fashion to an opaque pixel as described in Fig. 1.

The second issue (2) concerns that of an inelastic scattering event within the sample beam or reference beam destroying the coherence between the two beams.[38] Inelastic scattering at the crystal coupler may be reduced through careful consideration of the material and geometry of the coupler. The total inelastic scattering cross-section for electrons with ~100 keV energy tends to increase with increasing atomic number (Z) of the scattering material.[36] We may thus suppress the probability for inelastic scattering by forming the coupler from a low-Z material such as diamond or silicon. Diamond additionally possesses zone axes with short d-spacing values, supporting reduced losses to higher order diffraction pathways as described in (1) above. We may also decrease the total probability for inelastic scattering by reducing the thickness of the crystal coupler. We note that the trend of reducing inelastic scattering by reducing the thickness may change for ultrathin (< 5 nm thick) crystals. Inelastic scattering mechanisms such as surface and volume plasmon excitations are suppressed in ultrathin crystals, and retardation and interband transitions may dominate the inelastic scattering process.[39] Nevertheless, the crystal thickness should be optimized to both reduce inelastic scattering and achieve the desired intensity splitting between the sample and reference beams. An intensity splitting ratio of 99:1 between the reference beam and the sample beam would necessitate 16 circulations in the resonator to achieve maximum intensity transfer between the reference and sample states according to the Howie-Whelan equations, assuming a value of zero for the crystal-dependent parameter $s_{\text{eff}}$. If we consider a <220> oriented Si crystal splitter, and 100 keV electrons where $\xi_g$ = 75.7 nm, the Howie-Whelan equations described above allow us to calculate a suitable crystal thickness of 2.4 nm. Theoretically Si crystals with thicknesses of $t = n \times$ (75.7 nm) + 2.4 nm, where $n$ is an integer, could also achieve a similar splitting ratio. However, thicker crystals would lead to intensity loss, especially as the electron beam would have to pass through the crystal multiple times. (3) Correctly mapping the phase of the reference and sample beams back on to the crystal coupler after each circulation in the resonator is equivalent to positioning the interference pattern of the two beams with atomic resolution on the crystal lattice of the crystal coupler. As a result, it is clear that the resonator should be installed in a tool capable of producing atomically resolved images of a crystal lattice *i.e.* a high-resolution TEM. The high-resolution demands of re-imaging the sample and reference beams on the coupler with the correct relative phase would be reduced if an artificial crystal could be produced with a typical one-dimensional unit cell of say 10 nm. Although fabricating such a structure is not impossible with modern techniques in multilayer growth or lithography, it may introduce other issues because of atomic scattering. More research in this direction is needed.

The resonator may be formed using either a storage ring, or by using mirrors to create an electron cavity. We will consider the storage-ring design here, as the thin crystal requires high-energy electrons to maximize the mean-free-path of the electron traversing the crystal. A possible benefit of the storage ring design is that it can be combined with an electron energy-loss spectrometer (EELS) in a STEM to make use of the information in the inelastic channel. Here, we assume operation in the STEM mode to keep the description as simple as possible; however, the principle could also work in TEM mode, where many pixels are imaged simultaneously. A description of a system employing electron mirrors to create a free-space electron cavity is described in the next section.

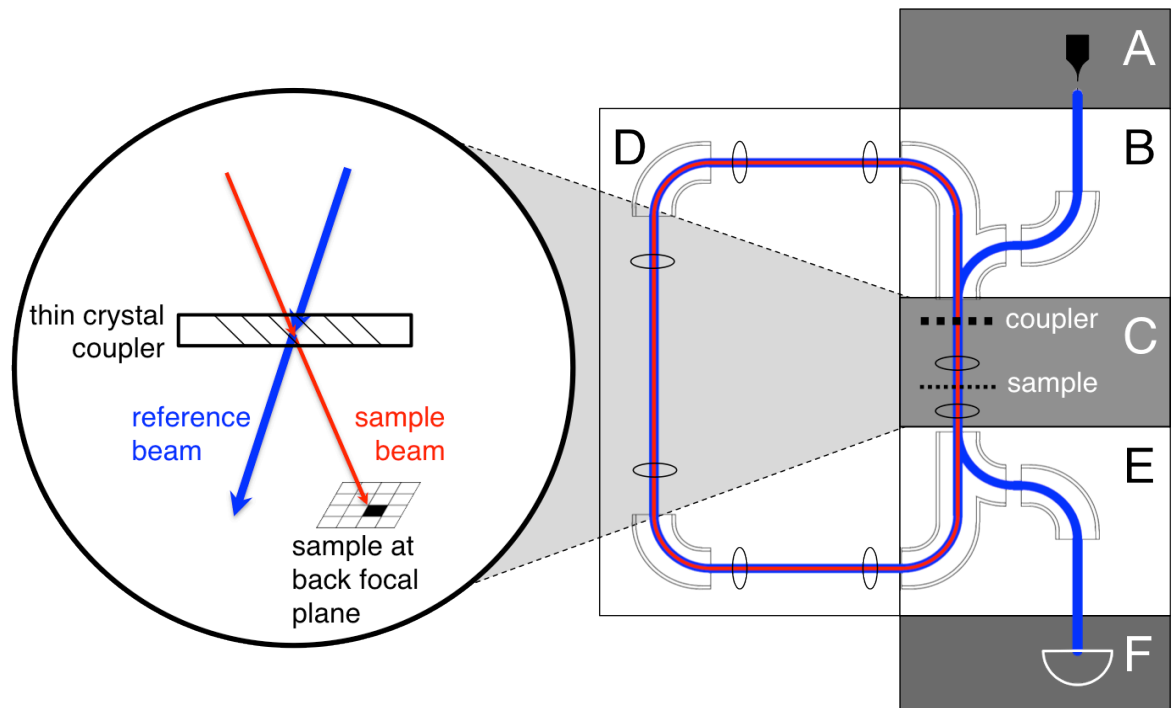

**Fig. 6.** Schematic drawing of the QEM system based on a thin crystalline coupler. (A) Electron source and beam forming optics. (B) Fast deflector allows insertion of electron pulse into the resonator. (C) Electron pulse is split at the coupler and a fraction of the electron-beam is focused onto the sample. (D) The sample and reference beams then propagate within the ring resonator and are projected back onto the coupler. (E) After *N* circulations in the resonator the electron pulse is deflected out of the resonator. (F) Measurement of the intensity in the reference beam at a detector will then establish whether the pixel in the path of the sample beam is transparent or opaque.

The layout of the electron optics required for the storage ring is shown schematically in Fig. 6. Briefly, the electron beam is formed in module [A], which consists of a high-brightness electron source such as that commonly used in a high-resolution (S)TEM. The beam is produced using an electron-optics system consisting of alignment optics, condenser lenses, and apertures that select a sufficiently small section of phase space to create a beam that is almost fully coherent. Module [B] contains a fast beam-deflection system enabling injection of an electron bunch into the ring-resonator. The deflection system should operate on a timescale significantly shorter than the circulation time of the electron bunch in the ring-resonator. The circumference of the resonator would be ~1 meter for typical STEM dimensions, thus giving a circulation time of approximately 5 ns for a 200 keV electron propagating at approximately $2 \times 10^8$ m s$^{-1}$. Upon entering the resonator the electron bunch is split at the crystal coupler such that one diffracted beam passes through the sample of interest. The sample is held in a conventional TEM sample holder residing in the immersion objective lens. If we assume that the electron beam is tightly focused at the sample plane (STEM-mode), this would suggest that the crystal coupler would be placed either, at the front focal plane, or at the back focal plane of the lens. In many microscopes, a diffraction aperture is placed in this position. Module [E] is a mirror copy of module [B] and acts to deflect the electron bunch out of the resonator to a detection system. This module may also contain magnification optics to further separate the direct beam from the diffracted beams and to allow normal sample imaging when the ring-resonator is deactivated. Module [D] contains bending magnets and lenses to recirculate the sample and reference beams back to the entrance plane of the coupler until the number of circulations required for intensity transfer between the reference and sample beams has been achieved.

One may wonder whether the addition of the resonator influences the resolution of the microscope or the longitudinal coherence of the beam. In the first pass of the beam through the sample, the resolution should be unaffected by the resonator, so the resolution of a modern STEM should be obtained. The STEM could even include an aberration corrector in section [C]. The STEM objective lens usually demagnifies the aberrations of the condenser lenses such that they may be ignored. Similarly, we assume that the aberrations of the resonator elements will also be demagnified. However, after each pass, the aberrations $C_o$ of the objective lens are added linearly to a value $NC_o$, while the resolution is proportional the square root of the total chromatic aberration ($\sqrt{C_c}$) and the 4$^{th}$ power root of the spherical aberration ($C_s^{1/4}$). Thus, to achieve atomic resolution, an aberration corrector will probably be required. One possibility is to insert a chromatic and spherical aberration corrector in the resonator by using the properties of the non-rotational symmetry of the deflectors.

If a crystal coupler is used, the exit wave of the coupler must be imaged onto the top surface of the crystal with sub-lattice resolution and accuracy. This resolution is easily obtained because the aperture angle of the beams at the coupler is small. The accuracy can be obtained by using deflection and rotation optical elements. The mechanical and electronic drift of the interference pattern on the coupler should be much less than the lattice spacing during the time of N circulations (~500 ns). Chromatic dispersion of the individual deflectors in the resonator should be mutually compensated.

If the reference beam and the sample beam travel over different paths through the resonator, there is a risk of losing the ability to interfere because of a relative longitudinal shift of the wave-packets. This effect has not yet been analyzed in detail, but we assume that it can be either eliminated or compensated. The same applies for the possible collapse through synchrotron radiation in the deflectors. On the latter effect, let us note that only radiation with sufficiently short wavelength to distinguish between the paths of the reference beam and the sample beam will lead to mutual decoherence.

We conclude that a QEM based on a "storage ring design" would be a complicated instrument, requiring many novel design features. However, we have not found reasons why it would be impossible to achieve atomic resolution in a quantum electron microscope with this design.

**2. Design based on diffractive electron mirror.**

As we have seen in the previous section a diffractive element such as a thin crystal may be employed as a coupler for a QEM design. One drawback of the crystalline beam splitter is that the electron beam has to pass through the crystal for diffraction to occur. An electron beam traversing a thin crystal will experience inelastic scattering resulting in a loss of signal. Consequently, we propose to diffract an electron beam at a periodic surface potential generated by applying a high voltage to an electrode on which a topographical grating with a nanometer to micrometer scale period has been fabricated (Fig. 7). Since the potential applied to the grating will be at or above the energy of the incident electron, the electron will reflect/diffract above the surface of the electrode and will not directly interact with the solid surface of the mirror. This approach thus eliminates the probability for inelastic scattering at the coupler. We call this device a grating mirror.

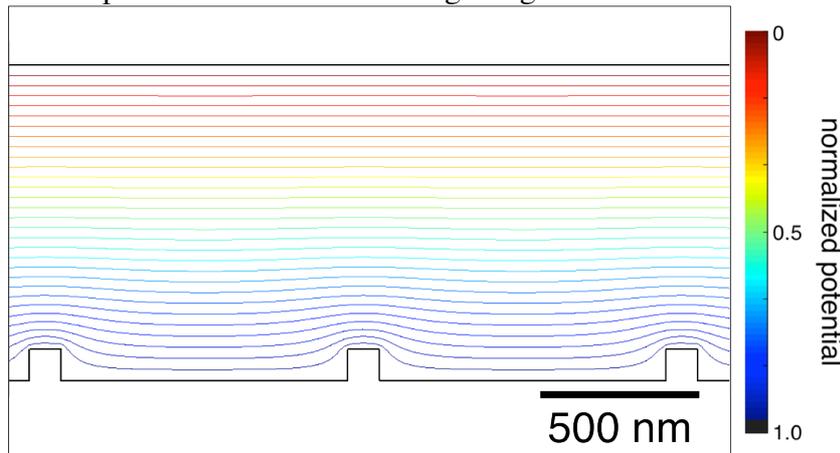

**Fig. 7.** A plot of a cross-section through the potential surface simulated at a 1-µm-period grating to which a static field of 30 kV/mm has been applied. The grating mesas have a square cross-section 100 nm in width and height. The potential difference between each of the equipotential lines shown in the plot is 1 V. The static field is generated by a bias potential applied between the grating surface and a planar, grounded counter electrode parallel to, and above the grating surface. A plane wave, such as that present in a collimated electron-beam, incident on such a potential surface would diffract according to the grating equation producing a linear array of diffraction spots in the back-focal plane of the collimating lens.

We consider diffraction from a sinusoidal grating mirror as equivalent to diffraction at a transmission phase grating, where diffraction is a result of the position-dependent phase change of the electron wave. A weak phase variation of $\Delta\phi$ (<<1) radians yields an amplitude in the diffraction plane with a delta-function on axis and the Fourier transform of the phase variation in the rest of the plane. So if the phase variation is a periodic function, the amplitude distribution also consists of delta functions with amplitudes of the "diffracted beams" of the order of $\Delta\phi$ in magnitude.[40] In the case of reflection from a grating, the position-dependent phase in the beam is introduced by the different path lengths of the trajectories upon reflecting from the surface. In a first approximation, we treat the whole electron mirror (decelerating

field – reflection – accelerating field) as a pure phase grating. This means that we assume that when the diffracted beams exit from the accelerating field, they have not yet interfered with the reference beam to create amplitude contrast. So at the exit from the mirror there is only a periodic phase variation on the exit wave. In the far field, the intensity distribution is then given by the Fourier transform of the exit waveform. This approximation should be accurate if the length over which the mirror field acts is short compared to the length scale of the instrument. This approach circumvents the difficulty of applying diffraction theory in a situation where the wavelength becomes infinite, which occurs at the position of reflection. Now the diffraction angle may be described by the grating equation.

$$n\lambda_{\text{electron}} = d \sin \theta \tag{1}$$

Here, $n$ is the diffractive order, $\lambda_{\text{electron}}$ is the de Broglie wavelength of the electron at the exit of the mirror, $d$ is the period of the grating and $\theta$ is the angle of diffraction. The de Broglie wavelength of a 5 keV electron is 17.4 pm, which results in a first-order diffraction angle of 17.4 μrad for a grating with a period of ~1μm. It is apparent from the grating equation that a higher energy electron beam or a larger grating period will result in a smaller angle of diffraction and thus lower spatial separation of the sample and reference beams in the resonator.

For small phase changes, we expect that we can apply a local WKB approximation since the changes are small on the scale of the wavelength. The phase change in going through the mirror is then:

$$\Delta\phi(x) = \int \frac{2\pi}{\lambda(x,z)} dz \text{ with } \lambda = \frac{h}{\sqrt{2meU(x,z)}} \tag{2}$$

Here, the integral is taken over the path of the electron. Both the λ(z) and the length of the path of the electron are x-dependent. The potential U in the grating mirror is a solution of the Laplace equation with a periodic boundary condition;

$$U(x,z) = E_z \cdot z + \Delta U \cdot e^{\frac{-2\pi z}{d}} \cdot \cos\left(\frac{2\pi x}{d}\right) \tag{3}$$

In equation 3 above, $d$ is the period of the sinusoidal potential variation with amplitude $\Delta U$ at $z = 0$ and $E_z$ is the mirror field. For a topographical grating with amplitude $A$, the surface acts as an equipotential surface. At some distance from the surface, the effect can be translated to a potential variation $\Delta U = AE_z$. Unfortunately, the integral for the phase change when performed for this potential distribution does not have an analytical solution. Nonetheless, one can see that it is possible to choose for reflection to occur at an equipotential surface closer or further from the surface by tuning the potential of the mirror. With that, the intensity in the diffracted beams is tuned. In fact, preliminary numerical calculations show that the diffracted intensity is very sensitive to the choice of reflection surface, which leads to the undesired effect of an electron-energy-dependent diffraction intensity. We have found that the grating period and potential gradient at the mirror surface must be large to minimize the effect of electron energy on diffraction intensity.

Fig. 8 shows a schematic drawing of a design for a microscope with a resonator including the grating mirror. A barn door (see next section for a detailed description) opens to allow an electron bunch to enter the resonator and subsequently closes to act as a flat mirror. A lens system focuses the reference beam through a hole in the sample. A second lens system creates a parallel beam onto the grating mirror. The reflected and diffracted beams are focused back on the plane of the sample. One first-order diffracted beam is focused on the sample. Higher order diffracted beams are blocked at this plane. During multiple passes through the resonator, the amplitude in the sample beam slowly increases. Effectively only half of the sample beam traverses the sample. The sample is illuminated both from the top and from the bottom. After $N$ cycles, the barn door is opened again and the bunch leaves the resonator. Between the electron source and the barn door is a beam separator that deflects the exiting electron bunch to a detector. Several sets of deflectors will be necessary to steer the beam correctly through the resonator. It is clear from Fig. 8 and the grating equation that a shorter grating period and lower electron energy give a larger separation of the reference beam and the sample beam and thus are preferential for ease of access to the sample beam. However, as the grating period is reduced, so too is the amplitude of the surface potential modulation at the mirror. A reduced modulation of the potential will ultimately lead to a reduction in diffraction efficiency. Likewise, a lower-energy electron is more sensitive to stray fields in the resonator resulting in poorer imaging resolution. However, these are only considerations for the very simple system shown in Fig. 8. The important parameter for a more complex system containing more lenses is the number of grating periods illuminated by the beam. A larger number of illuminated grating periods results in a sharper diffraction spot and thus improved resolution in the diffracted beam.

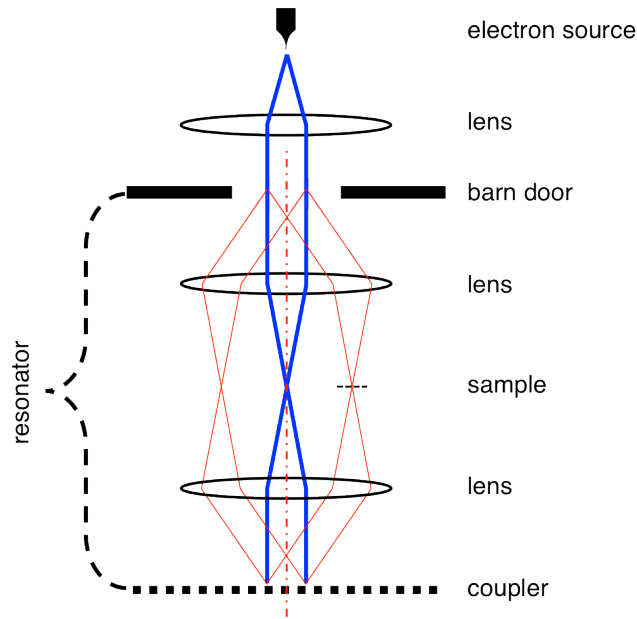

**Fig. 8.** Schematic drawing of the QEM system based on a nanostructured electron mirror coupler. The thick (blue) rays represent the reference electron-beam while the thin (red) rays represent the sample electron-beam. The sample is inserted into the path of one of the red rays at the crossover near the center of the resonator. The coupler at the base of the resonator is a diffractive mirror such as that shown in Fig. 7.

An example layout of a more complex microscope using a grating mirror is shown schematically in Fig. 9. The electron beam is formed in a module (A) which consists of a high-brightness electron source such as that commonly used in a high-resolution (S)TEM. The beam is first limited by apertures, which select a sufficiently small section of phase-space, to create a beam that is almost fully coherent. As the electron energy is still low, this region is a suitable position for a barn door that upon closure acts as an electron mirror. The beam is now accelerated and enters a column similar to that in a conventional (S)TEM with condenser lenses, alignment deflectors, stigmators, a high-resolution objective lens and the sample holder (B,C). After the objective lens and a few magnification lenses (D), there is a module in which the grating mirror is held at the potential of the electron source (E). In this module, the beam is decelerated towards the grating mirror. A set of electrostatic deflectors allows the beam to be deflected towards a hole in the mirror so that, after the required number of bounces, the beam can exit the resonator. The beam is accelerated again into the last module of the microscope. This module contains the detector and may also contain magnification optics to further separate the direct-beam from the diffracted beams, and to allow normal sample imaging when the resonator is deactivated.

The analysis of lens aberrations for this design is similar to that discussed for the ring-resonator, so an aberration corrector would also be very useful here. An interesting opportunity arises in this system because we use mirrors located in a conjugate plane of the objective lens' back-focal plane. A mirror at that position can be used as aberration corrector, both for spherical and chromatic aberration.

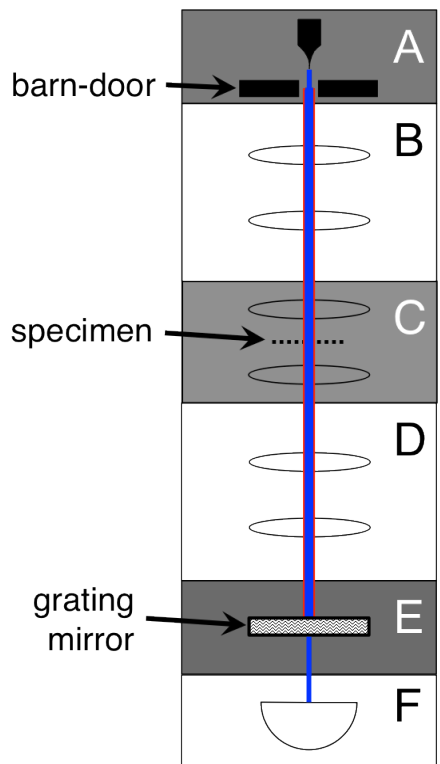

**Fig. 9.** Schematic of a high-resolution QEM system based on a grating mirror coupler. (A) Electron source, entrance barn door and mirror (at high voltage). (B) Condenser lenses. (C) High-resolution immersion objective lens. (D) Magnification lenses. (E) Grating mirror and exit barn door (at high voltage). (F) Measurement of the intensity in the reference beam at a detector will then establish whether the pixel in the path of the sample beam is transparent or opaque.

In conclusion, a design of a QEM with a resonator between two mirrors seems entirely feasible and considerably simpler than a design with a ring-resonator. However, it relies on the as yet unproven operation of the "grating mirror".

### 3. Design based on a coupler using a standing wave of light.

A coherent beamsplitter can also be realized via the diffraction of electrons at a standing wave of light. This technique was originally proposed by Kapitza and Dirac in 1933[41] and was later used to diffract electrons,[42] atoms[43] and molecules.[44] A review on the Kapitza-Dirac effect in electron matter wave optics has been prepared by Batelaan.[45]

A beam splitter made of light offers several advantages, most notably that there is no absorption and negligible inelastic scattering of electrons. The required optics can be mounted far from the electron beam such that charging and patch potentials do not pose a serious problem. Also, as we will see in the following, the diffractive power of such a beamsplitter can simply be tuned by varying the power of the laser and it would thus be possible to alter the number of bounces required for a QEM measurement.

The Kapitza-Dirac effect can be described as the diffraction of electrons at the ponderomotive potential $U_\text{p}$[46] of the standing light wave: $U_\text{p} = \frac{e^2 I}{2m\epsilon_0 c\omega^2} \cos^2(kx)$, where $k = \frac{2\pi}{\lambda}$, $\omega = \frac{2\pi c}{\lambda}$ and $I$ is the laser intensity. This scattering can also be understood in a particle picture, where absorption of a photon with momentum $\hbar k$ and stimulated emission induced by a photon with momentum $-\hbar k$ leads to a total momentum transfer of $2\hbar k$ and a corresponding deflection of the electron beam by an angle $\Theta = \frac{2\hbar k}{p_z}$, where $p_z$ is the longitudinal momentum of the electron.

A beamsplitter for a QEM based on a standing light wave could be realized in two different regimes (see Fig. 10): For a strong ponderomotive potential ($U_\text{p} \gg \frac{\hbar^2 k^2}{2m_\text{e}}$), in the so called diffractive regime, this will lead to an interference pattern showing multiple diffraction peaks,[42] each corresponding to a transverse momentum transfer $\Delta p_x = \pm 2n\hbar k$, $n \in \mathbb{Z}$. The probability of finding an electron in the $n^\text{th}$ diffraction peak will be given by $I_n = J_n^2\left(\frac{U_0 t}{\hbar}\right)$, where $J_n$ is the $n^\text{th}$ Bessel function of the first kind.

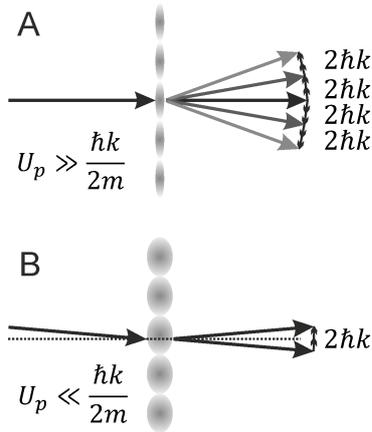

**Fig. 10.** A standing light wave as a beam splitter in a QEM could either work in the diffractive regime (A) or in the Bragg regime (B). The interference orders are separated by $2\hbar k$ in momentum space.

In the Bragg regime ($U_\text{p} \ll \frac{\hbar^2 k^2}{2m_\text{e}}$), only two diffraction orders couple[47] and they are again separated in momentum space by $\Delta p_x = 2\hbar k$. The probability of finding an electron in each of the two diffraction orders will follow a *Pendellösung* with $P_{\hbar k} = \cos^2\left(\frac{U_\text{p} t}{4\hbar}\right)$ and $P_{-\hbar k} = \sin^2\left(\frac{U_\text{p} t}{4\hbar}\right)$ analogous to the solution obtained in the original QEM proposal.[25] Note that in a QEM there would be $N$ interactions between an electron and the standing light wave and the argument in the above solutions would have to be replaced by $\frac{N U_\text{p} t}{4\hbar}$. Also, the electrons have to encounter the standing light wave at an angle $\Theta = \pm\frac{\hbar k}{p_z}$ in order to fulfill energy and momentum conservation.

In the diffractive regime, the setup would look like the one sketched in Fig. 6, with the standing light wave replacing the thin-crystal beam splitter. Again, all higher diffraction orders would have to be blocked in order to avoid losses to the higher diffraction orders as discussed in section 1 describing the design based on a thin-crystal coupler. One possible electron optical design for a beam splitter operated in the Bragg regime is shown in Fig. 11.

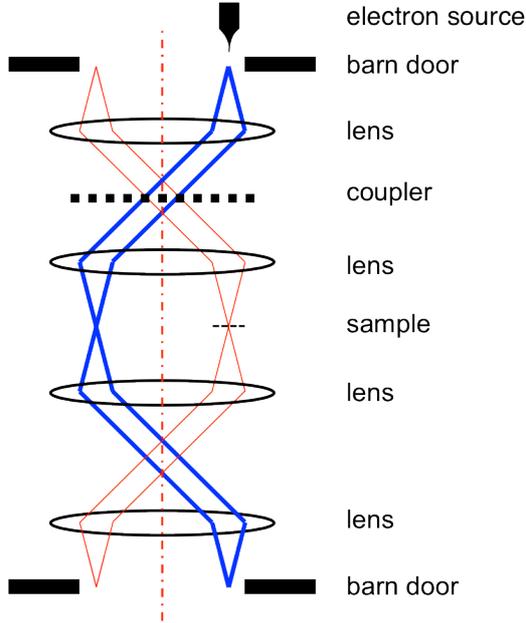

**Fig. 11.** Conceptual design of a QEM with a standing light wave as the coupler element. The electron beam enters the resonator off-axis, thus enabling the formation of a two-beam system.

It is interesting to consider the laser intensity requirement for such a beam splitter. In the Bragg regime, the wavefunction of an electron is completely transferred from the reference beam to the (empty) sample beam if $\frac{U_0 t}{4\hbar} = \frac{\pi}{2}$. In the diffractive regime, a similar statement can be made as the zeroth order Bessel function has its first zero at $\frac{U_0 t}{\hbar} \approx 2.4$. For an electron passing through a focused Gaussian laser-beam of waist $r_0$, we get $\int_{-\infty}^{\infty} U_p(t) dt \approx \frac{U_p r_0 \sqrt{\frac{\pi}{2}}}{v}$, where $v = \sqrt{\frac{2E_{\text{kin}}}{m}}$ is the velocity of the electron. With $I = \frac{2P}{\pi r_0^2}$, where $P$ is the total power of the continuous wave laser, and after $N$ interactions our laser power requirement can be written as in equation 4.

$$P = 4\omega^2 r_0 \sqrt{E_{\text{kin}} m \pi^3} \frac{\epsilon_0 c \hbar}{N e^2} \qquad (4)$$

It is thus tempting to choose a long laser wavelength, however this would decrease the diffraction angle. Lower electron energies on the other hand will both decrease the power requirement and increase the de Broglie wavelength of the electrons. For $\lambda = 532$ nm, $E_{\text{kin}} = 1$ keV and $N = 100$ we get $P \approx 3.7$ kW for a focal waist of $r_0 = 10$ μm or $P \approx 370$ kW for a focal waist of $r_0 = 1$ mm (the transition from the diffractive to the Bragg regime $U_p = \frac{\hbar^2 k^2}{2m_e}$ is reached for $r_0 = 116$ μm).

A standing light wave of this intensity can either be realized with two counter propagating laser pulses or by enhancing a continuous wave laser in a cavity. In the Bragg regime, it might be especially convenient to relax this laser power requirement by working with an elliptical beam profile that is compressed perpendicular to the direction of the electron beam.

**4. Design based on a double potential well coupler.**
The motion of slow electrons with kinetic energy on the order of several electronvolts can be precisely manipulated in free space using high-frequency electric fields. This capability allows the fabrication of 2-d traps for electrons based on the generation of the necessary microwave fields by means of properly shaped electrodes on a planar substrate.[48,49] The generation of microwave fields by means of a planar microwave chip provides ease of scalability and the flexibility to engineer versatile guiding potentials in the near-field of the microwave excitation. This feature makes surface-electrode structures ideally suited for the implementation of a double-well potential as originally proposed by Putnam and Yanik.[25]

The approach described here is based on a similar principle as the Paul trap,[50] which is widely used to trap and guide the motion of atomic ions.[51] In a 2-d Paul trap, the transverse confinement of the guide relies on the formation of a time-averaged harmonic pseudopotential resulting from a high-frequency electric quadrupole potential that oscillates in the gigahertz frequency range. Stable electron confinement requires that the potential gradient experienced by the electron is nearly constant over the amplitude of the electron's oscillation. A double-well pseudopotential can be generated based on the same principles using a microwave electric potential with hexapole instead of quadrupole symmetry. The hexapole symmetry is the lowest-order multipole that supports two intersecting potential minima and hence creates a beam splitter.[52]

The demonstration of an electron beam splitter is based on the guiding techniques described above,[53] which will enable the coupling unit in this QEM approach. Fig. 12(a) shows the electrode layout of the planar beam splitter. The electric field above the guiding chip can be transformed from quadrupole to hexapole symmetry along the horizontal y-direction by means of a tapered central electrode. The hexapole symmetry of the electric field leads to the creation of a double-well pseudopotential with two minima, separated by a potential barrier. The simulated pseudopotential is plotted in Fig. 12(b-d), showing the transition from a single to a double well. The separation of the double well can be controlled by adjusting the width of the central electrode. Furthermore, we have numerically optimized the shape of the chip electrodes to provide a guiding potential that allows smooth beam splitting, producing two spatially separated electron beams.

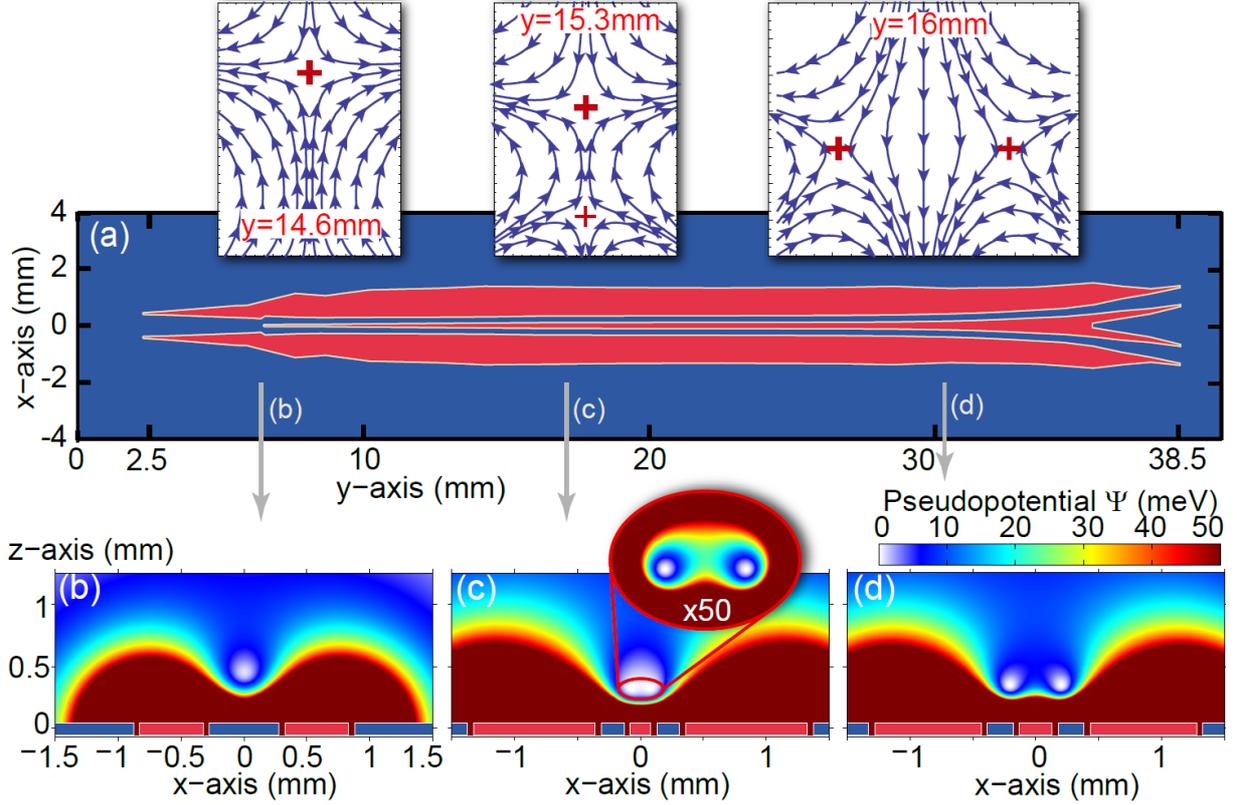

**Fig. 12.** Electrode design of the planar electron beam splitter and pseudopotential simulations. (a) The numerically optimized layout of the microwave driven electrodes is shown in red. The blue areas are electrically grounded. The insets show the electric field generated above the surface at three selected positions. (b) Cross-section normal to the electrode plane at $y = 6.5$ mm. A single pseudopotential minimum forms at a height of 450 $\mu$m above the substrate surface providing harmonic confinement. (c) At $y = 17$mm the additional central electrode, with a width of 160 $\mu$m, creates a double-well pseudopotential with a separation of 150 $\mu$m between the minima. A four-fold magnified image is shown in the inset with a 50 times amplified color code. By increasing the width of the center electrode the separation of the double-well minima is gradually increased. (d) At $y = 30$ mm the central electrode is 260 $\mu$m wide, leading to a separation of the minima of 400 $\mu$m.

However, to implement the scheme proposed by Putnam and Yanik using such a microwave guide with two arms, a few challenges have to be overcome. A single electron has to be coupled into one of the arms, which is then supposed to provide a tight confinement and at the same time enable transfer of electron wave amplitude into the second arm. To realize these two conflicting requirements (tight confinement and coupling), we present a modified scheme where the coupling potential between the two arms is not constant. Thus our proposed electron guide is divided into two sections, one of which enables coupling between the resonator arms, while the other realizes a tight confinement and separates the two arms to allow encounters with the sample.

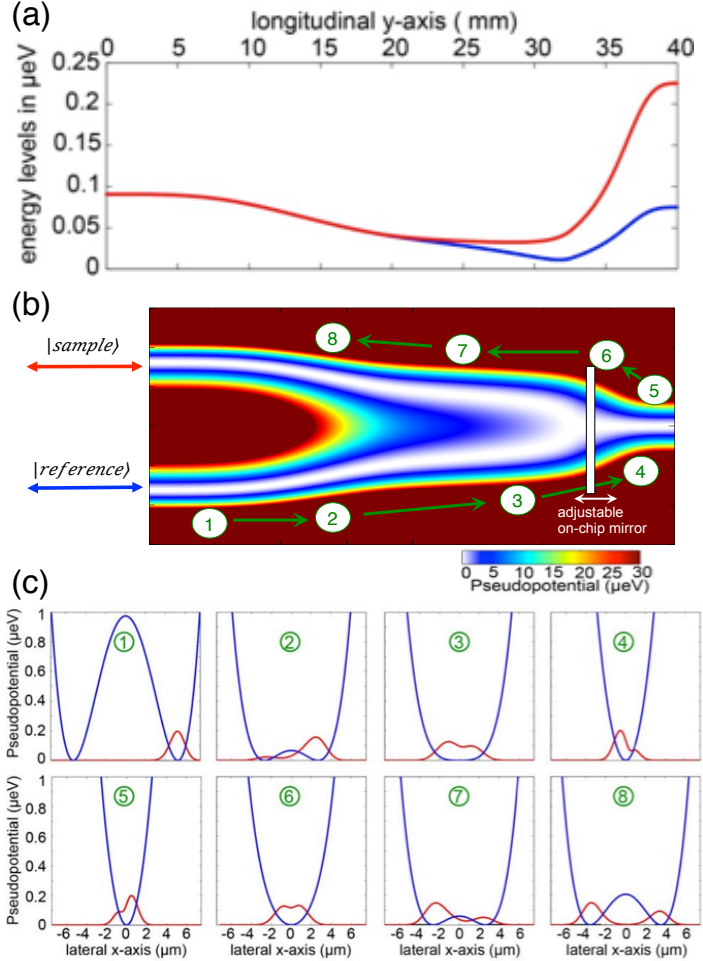

**Fig. 13.** Chip-based guided electron-beam coupler. (a) Energy level of the two lowest quantum states of the confining potential. (b) Numerically optimized beam splitter potential as coupler in the QEM scheme. The electron optics required to focus and reflect the electron beam are shown in Fig. 14. See text for further explanation. (c) Evolution of a wave-packet in the splitting potential. The blue curve shows the potential in transverse direction. The absolute value of an electron wavefunction is plotted in red.

Fig. 14 shows a potential implementation of the QEM scheme using the planar beam splitter as a coupling element between two electron resonators. The envisioned scheme uses electron wave packets that are laser-triggered from a metal nano-tip as a coherent source.[54,55] The electrons are coupled into the transverse ground state of the lower arm of the beam splitter potential (Fig. 13(b)). As described above, this localized |*reference*⟩ state is actually a 50/50 superposition of the two lowest quantum states of the double well potential.

Fig. 13(a) shows the energy levels of the two lowest states along the beam splitter chip. As the electron passes the junction of the potential at $y \approx 35$ mm (Point 4 in Fig. 13(b,c)), the confining potential transforms from a double-well to a single-well transverse guiding potential. As a result, the electron, which is originally in the ground state of the lower arm, now occupies a superposition of the two lowest energy states of the single-well potential. This superposition state alternates temporally between the two lowest eigenstates. The frequency at which the electron's state oscillates is proportional to the energy difference of the respective states and may reach a frequency of up to $\omega \approx 2\pi \cdot 58$ MHz. At $E_{kin} = 1$ eV, one period of this frequency corresponds to 10 cm of travel for the electron; however, the electron may be slowed to arbitrarily low energies while confined above the chip.

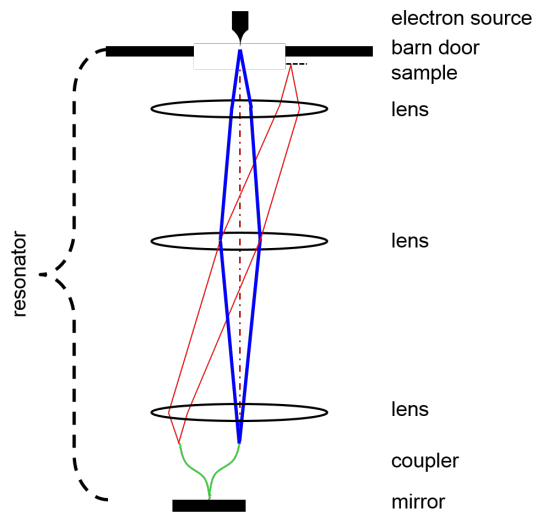

**Fig. 14.** Schematic of a resonator system using the planar, on-chip, double-well beamsplitter (green line centered above mirror). The chip acts as a coupling element between the two arms of the electron resonator. The reference beam (thick blue line) enters the resonator through the "barn door" and is focused to the right arm of the coupler (lower arm in Fig. 13(b)). The coupler reflects the reference beam and transfers a small part of the amplitude into the sample beam (thin red line), which exits the left arm. Reference and sample beam are focused by the same electron optics onto the (now closed) "barn door" plane, which also serves as the sample plane in this scheme. Other optical designs where the sample is in field free space are also possible.

When the electron reaches the end of the chip at $y = 40$ mm, it is reflected by an electron mirror and driven back to the double-well section of the guide, where the two minima are well separated. The superposition of the first two eigenstates of a single well is thus turned into a superposition of the first two eigenstates of a double-well, corresponding to a splitting of amplitudes between the wells. Part of the amplitude from the $|reference\rangle$ state has been transferred to the $|sample\rangle$ state. How the amplitude splits between the two wells depends on the amount of time the electron remains in the single-well state, and hence the evolution of the superposition state during this time. Fig. 13(c) illustrates the subsequent evolution of the wave-packet at several positions along the guide. The part of the wave-packet that finally ends up in the upper part of the potential ($|sample\rangle$ state) is ejected from the guide, accelerated, and then focused onto the specimen or, if the specimen is transparent or absent, reflected back by a second mirror to be re-injected in into the guide. The other part of the wave-packet following the lower part of the potential ($|reference\rangle$ state) is also ejected from the guide and reflected by the closed 'barn-door' without interaction with the specimen. By adjusting the point where the electron gets reflected inside the coupler, the time spent in the single well and thus the amplitude transfer from the $|reference\rangle$ to the $|sample\rangle$ state can be adjusted. This control enables tuning the success probability of the quantum measurement and thus the damage to the specimen. All the involved electron mirrors can be realized by electrostatic potentials. The "barn door" at the entrance to the resonator can be realized with schemes as discussed in the preceding section.

Notes on the barn door

One challenge in building an electron resonator is coupling the electron into the resonator, and then out for detection. This can be done by either lowering one of the defining potentials of the resonator at the moment the electron needs to enter/exit the resonator structure (see Fig. 15(a)), or by quickly decreasing/increasing the kinetic energy of the electron (see Fig. 15(b)), or by a pulsed deflection of the electron beam (see Fig. 15(c)). All approaches require this to be done on a timescale that is short compared to the round-trip time of an electron in the resonator. Additionally, it is important to consider whether the electron beam should enter the resonator as a focused beam or as a broad collimated beam. Our preliminary designs consider a focused beam entering the resonator, thus allowing all electrons to remain in a paraxial condition during barn-door operation. Consequently, these paraxial electrons will be less susceptible to the strong aberrations expected far from the optical axis of the barn-door.

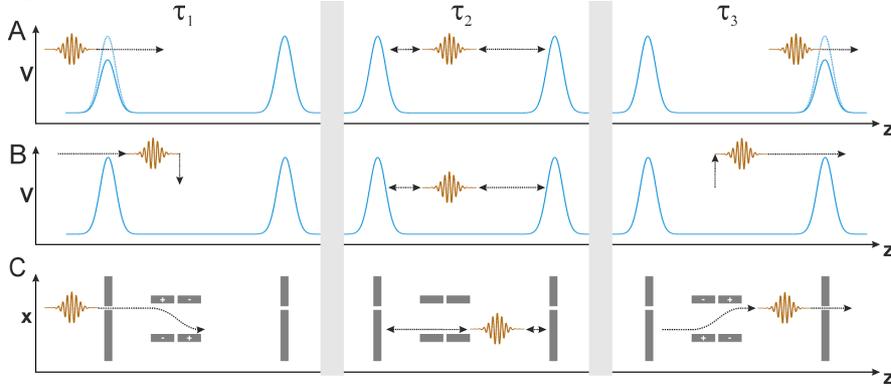

**Fig. 15.** Three concepts for coupling electrons in and out of the resonator structure: (A) At time $\tau_1$ the potential $V(z,t)$ on the entry barn door is lowered, such that the kinetic energy of the electron is commensurate with entry to the resonator structure. After one reflection from the exit barn door, the potential on the entry barn door has to be raised sufficiently to reflect the electron. The electron can be out-coupled after multiple bounces at time $\tau_3$ by lowering the potential applied to the exit barn door. (B) Alternatively, the potentials on the resonator end caps can remain constant, if the kinetic energy of the electron is decreased sufficiently at $\tau_1$ and later increased again at $\tau_3$. (C) Pulsed lateral deflection constitutes another possibility to trap an electron within electrostatic resonator end caps.

The resonator round-trip time depends on the energy of the electrons and the geometry of the resonator, which depends on the period of the two-state coupler used for the specific design. We can perform a crude estimate for the design shown in Fig. 11 by assuming that at the sample plane, the distance between the reference beam and the sample beam should be at least $\Delta x = 5\,\mu$m, in order to have sufficient access to sample beam. At constant electron energy, the time for an electron to travel from the standing light wave to the sample plane will be $\tau > \frac{m_e \Delta x}{2\hbar k}$, which amounts to 1.8 ns, where $k = 2\pi/\lambda$, with $\lambda = 532$ nm. This is independent of the longitudinal velocity of the electrons. The total round-trip time depends on the number of lens elements and the potential-energy landscape within the resonator, but will be on the order of 10 to 30 ns for the resonator design shown in Fig. 11. This suggests that coupling of electrons into and out of the resonator should be controlled on a timescale of about 1 ns.

While pulsed electrostatic ion traps have been realized with gating times on the order of hundreds of nanoseconds,[56] gating times on the order of one nanosecond will require careful design and impedance matching of the electrodes.

For a two-state coupler based on a thin crystal, the round-trip time could potentially be shorter due to the larger diffraction angle. Sub-nanosecond in-coupling and out-coupling of electrons might be realized using precisely timed THz[57] or optical control[58] of electron trajectories and energies.

**Discussion**

Acquiring atomic resolution, 3-d images of biological specimens and macromolecules with negligible damage to the sample is presently one of the greatest challenges in electron microscopy. One approach toward meeting that challenge could be via the quantum principle of IFM. That principle has been realized already for optical systems, but this paper contains the first sketches of modified electron microscopes in which quantum interrogations or an IFM could be performed at atomic resolution. However, we note that the principle of interaction-free imaging has thus far only been shown to be capable of generating binary, black and white images. This imaging mechanism can be applied to high-contrast samples consisting of pixels that are not fully opaque or transparent. However, true grayscale imaging using a QEM is expected to result in specimen-damage comparable to that in normal electron microscopy, an issue that was discussed recently in another paper from members of our collaboration.[27] If we consider real samples, this begs us to question how transparency is defined in electron microscopy. Samples can be opaque due to an interaction with the sample that destroys the coherence between the reference and the sample beam, by large angle incoherent scattering for example. In addition to scattering, real specimens also induce phase-shifts to the electron wave, which can alter the interference pattern produced at the coupler. In fact, it turns out that a phase-shift results in a similar effect to an opaque pixel in a QEM if the number of round-trips is high. However, if the phase-shift is known or can be calculated, a commensurate phase-shift could intentionally be added to the reference beam to compensate for the phase-shift caused by the specimen. Consequently, there may be potential for the development of additional operating modes in the QEM system.

The designs of microscopes for interaction-free imaging presented here are based on different ideas for a "two-state-coupler": a device that can bring electrons controllably from one state into another state and back again by repeatedly passing the electrons through it. The four different types of couplers considered in this work have been proven in principle to operate as electron beam-splitters for coherent electron beams, however, their use to recombine spatially separated electron beams remains to be demonstrated. Consequently, the present discussion is intended to help direct the development of such couplers. Note that we have considered using an electron biprism as a coupler, since such a device is able to direct part of a coherent electron wave function into a separate beam, however we concluded that a biprism is not able to slowly add more amplitude to a sample beam, which is a requirement for the desired coupler. One of the problems we foresee with the two-state couplers is that more than one (or two as we need in some designs) side beam is created, thus disturbing the principle. However, it may be expected that there is little amplitude in these extra beams after one single interaction with the coupler and thus, if these extra beams are blocked, very little intensity will be lost from the experiment during each round-trip in the resonator. The effect of small losses on such a measurement has been discussed in a previous report by Kwiat *et al.*[17]

Another untested idea in the designs presented here is the possibility of creating an "electron resonator". Synchrotrons and other particle accelerators prove that the injection of a particle beam into a device that holds that particle for a long time is possible; however, a design for such a device that is compatible with present day electron microscopes is still outstanding. It is clear that a compatible resonator device will require a fast switching element to support both in-coupling and out-coupling of the electron-beam to and from the device. The design of a fast switching barn-door described here has been simulated and it seems that the aberrations can be small enough to avoid disturbing the beam, but as we have stated a resonator has not yet been demonstrated.

If the proposed quantum electron microscopes are to achieve atomic resolution, it means that the focused electron probe should be on the order of Ångstroms in diameter and it should return to exactly the same position after each cycle through the resonator. In the case of a thin crystal coupler, there is a greater challenge in that the interference pattern of the reference beam and the sample beam must be aligned to the atomic lattice that produced diffraction. Even a fraction of a lattice spacing shift would disturb the slow build-up of intensity in the sample beam. Magnification, rotation and shift must also be correct.

The effect of lens aberrations on a cycling beam has been discussed already in the text. The likely consequence of needing to correct for these aberrations in the ultimate high-resolution instrument complicates the design further, although we are optimistic that the intrinsic availability of a mirror or a curved axis would make the correction of aberrations easier than in present day microscopes.

**Conclusion**

It is possible to design a "quantum electron microscope" in which quantum interrogations can be performed at atomic resolution. In the approach described in this paper, such a microscope primarily uses standard components common to present day microscopes, such as a high-brightness electron source, a high-resolution objective lens and specimen holder, transfer lenses, and a detector. New components that are required are: (1) a unit in which the electron beam can travel repeatedly through the sample, we call this a "resonator"; (2) a two-state coupler that splits the electron beam coherently into two parts and supports oscillatory intensity transfer between the two parts; (3) a unit that allows the electron beam to enter or exit the resonator, a "barn door". The localized two-state coupler is, for electrons, a novel device for which we describe four possible embodiments (a thin crystal, a grating mirror, a standing light wave and a an electro-dynamical pseudopotential), which at this time are being developed in parallel to determine the most practical option.

Before embarking on a project to build a full atomic-resolution quantum electron microscope, simplified versions of these designs can be used to test the novel components, the coupler, resonator and barn door. At the same time, further theoretical work is needed in order to understand the images that could be obtained through the principle of quantum interrogation and to quantitatively estimate the advantages for imaging biological samples that a quantum electron microscope could yield.

**Acknowledgements**

This research is funded by the Gordon and Betty Moore Foundation. The authors would like to thank M. F. Yanik, W. P. Putnam, J. Spence, F. Hasselbach, and D. van Dyke for helpful discussions.

**Author contributions**

KKB, MK, PH and PK conceived the initial microscope architectures described herein. CSK performed the experiment for Fig. 5. PK, RGH, YY, TJ, PW, ST, KKB and PH wrote the paper, coordinated by RGH. All authors contributed to discussions in the quantum electron microscopy project.